\documentclass{osa-article}
\journal{oe}
\newcommand{\blue}[1]{\textcolor{blue}{ #1}}

\usepackage{hyperref}
\begin{document}
\title{Polarisation-insensitive generation of vector modes using a digital micromirror device}

\author{Carmelo Rosales-Guzm\'an\authormark{1*}, Xiao-Bo Hu\authormark{1}, Adam Selyem\authormark{2}, Pedro Moreno-Acosta \authormark{3}, Sonja Franke-Arnold\authormark{4}, Ruben Ramos-Garcia \authormark{4} and Andrew Forbes\authormark{5}}

\address{\authormark{1}Wang Da-Heng Collaborative Innovation Center for Quantum manipulation \& Control, Harbin University of Science and Technology, Harbin 150080, China}
\address{\authormark{2}Fraunhofer Centre for Applied Photonics, G1 1RD, Glasgow, Scotland} 
\address{\authormark{3}Instituto Nacional de Astrof\'isica, \'Optica y Electr\'onica, Luis Enrique Erro 1, Tonantzintla, Puebla, M\'exico}
\address{\authormark{4}School of Physics and Astronomy, University of Glasgow, G12 8QQ, Glasgow, Scotland}

\address{\authormark{5}School of Physics, University of the Witwatersrand, Johannesburg 2050, South Africa}

\email{\authormark{*}carmelorosalesg@hrbust.edu.cn}


\begin{abstract}
In recent time there has been an increasing amount of interest in developing novel techniques for the generation of complex vector light beams. Amongst these, digital holography stands out as one of the most flexible and versatile with almost unlimited freedom to generate scalar and vector light beams with arbitrary polarisation distributions and spatial transverse profile. Recently, we put forward a novel method to quantify the non-separability of vector modes in which we reported first measurements of a compact and robust device to generate such vector modes that fully exploits the polarisation-independence of Digital Micromirror Devices (DMDs). In this manuscript we fully characterise this device and provide qualitative and quantitative analysis of the generated modes. First by reconstructing their transverse polarisation distribution, using stokes polarimetry, followed by a measure of their degree of non-separability, determined through the concurrence. 
\end{abstract}

%

\section{{Introduction}}
 
Complex vector light fields are fascinating states of light that have captured the interest of researchers across a wide variety of fields where they have found a myriad of applications \cite{Rosales2018Review}. In vector light fields the spatial and polarisation Degrees of Freedom (DoF) are coupled in a non-separable way, giving rise to a non-homogeneous distribution of polarisation that holds many interesting properties \cite{Otte2016,Galvez2015,Bauer2015,Beckley2010,Otte2018}. In addition, this non-separability has been identified as the classical analogue of local quantum entanglement, enabling quantum-like phenomena at the classical level \cite{Spreeuw1998,ChavezCerda2007,Qian2011,Aiello2015,konrad2019,toninelli2019concepts,forbes2019classically}. In the last decade several techniques have been proposed to generate vector beams, including interferometric arrays \cite{Tidwell1990,Niziev2006,Passilly2005,Mendoza-Hernandez2019}, liquid crystal wave plates\cite{Marrucci2006,Naidoo2016}, glass cones\cite{Radwell2016,Kozawa2005}, nanomaterials\cite{Devlin2017}, Spatial Light Modulators (SLMs)\cite{Davis2000,Maurer2007,Moreno2012,Mitchell2017,SPIEbook,Rosales2017,Rong2014,Liu2018} and more recently Digital Micromirror Devices 
(DMDs)\cite{Ren2015,Mitchell2016,Scholes2019,Gong2014}. Ultimately, most techniques aim for the full control over the phase, amplitude and polarisation of light towards the generation of vector beams with arbitrary polarisation distributions. In this way, computer-controlled devices, such as SLMs and DMDs stand out as some of the most flexible and versatile. Crucially, while SLMs are polarisation dependent, allowing only the modulation of linear polarisation (typically horizontal), DMDs can modulate any polarisation, a property that has gone almost unnoticed since common experimental setups are very similar to SLM-based ones. That is, in order to generate arbitrary vector beams with an SLM, the transverse profiles of both polarisation components have to be manipulated independently, either in interferometric arrays  containing one or two SLMs \cite{Maurer2007,Rosales2017,Liu2018,Mendoza-Hernandez2019,Galvez2012,Niziev2006}, or via a temporal sequence using a double pass over a single SLM \cite{Rong2014,Otte2018b}. Crucially, the polarisation-independence of DMDs allows for the generation of complex light fields in a more compact way, at high refresh rates ($\sim$30 KHz) and with monochromatic sources over a broad band of the visible spectrum \cite{DMDManual}. 

In recent time we reported a novel method to quantify the non-separability of vector modes in which we implemented a compact and robust device to generate such vector modes that fully exploits the polarisation-independence of DMDs \cite{Selyem2019}. In this manuscript we fully characterise this device and provide qualitative and quantitative analysis of the generated modes. Our device comprises the illumination of a DMD with two beams of orthogonal polarisation impinging at different angles to modulate the spatial degree of freedom of both polarisation components in a single pass. For this purpose, we display a multiplexed hologram consisting of two independent holograms with different spatial carrier frequencies on the DMD. The hologram is designed to overlap the first diffraction order of each incoming beam so that they propagate along the same optical axis. In this way, the optical mode after the DMD emerges as a structured light mode. Given that both constituent holograms are independent, our device can generate any vector light beam with tunable degrees of non-separability, and with arbitrary spatial and polarisation distributions. To show this, we generate vector beams with arbitrary polarisation distributions using the Laguerre-Gaussian spatial modes, so-called Poincar\'e beams, such as "spiders", "webs" or "lemons" \cite{Galvez2012}. We further generate vector beams with arbitrary polarisation distributions as defined on the High Order Poincar\'e Sphere (HOPS) \cite{Milione2011}. Finally, we perform an exhaustive evaluation of the generated modes to provide further details of their quality. First, we perform a qualitative measure based on a reconstruction of their transverse polarisation distribution, using Stokes polarimetry \cite{Zhaobo2019}, followed by a quantitative measure to determine the degree of non-separability between the spatial and polarisation DoF \cite{Ndagano2016,McLaren2015,Selyem2019}. All measurements indicate that vector modes generated with this polarisation-insensitive approach possess high quality.   

\section{Polarisation-insensitive generation of complex vector modes}

\begin{figure}[h!]
\centering
\includegraphics[width=\textwidth]{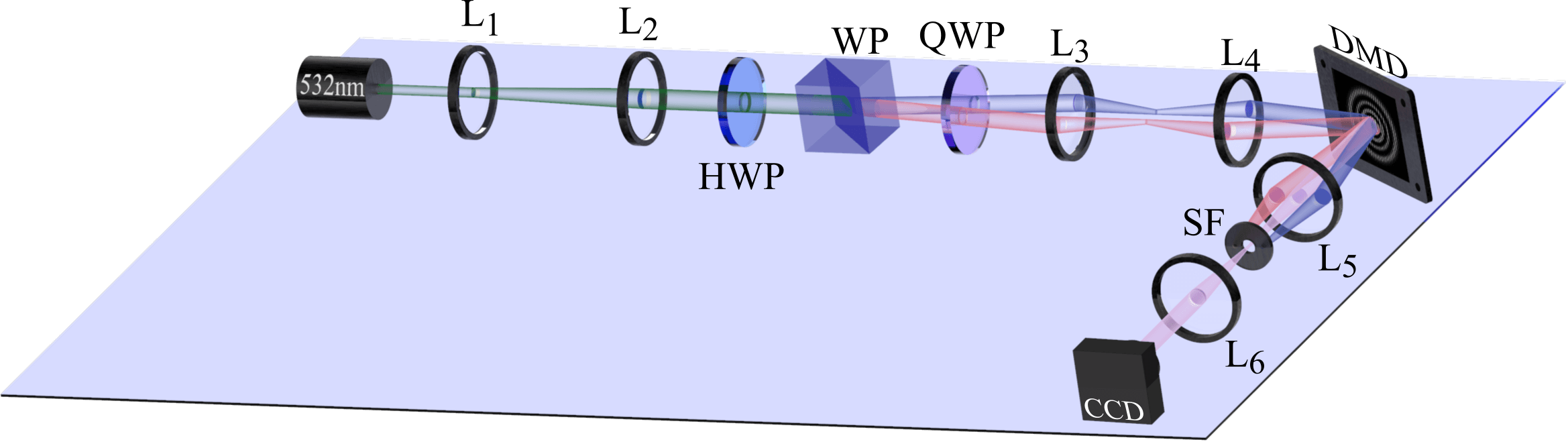}
\caption{Schematic representation of our polarisation-independent device for the generation of vector modes. A linearly polarised collimated and expanded beam is passed through a half wave-plate (HWP) to generate a diagonally polarised beam, which is then separated into its vertical and horizontal components using a Wollaston prism (WP). The desired vector mode is generated by overlapping these two orthogonally polarised beams in the correct diffraction orders of two multiplexed gratings displyed on the DMD. The desired diffraction order is selected with a Spatial Filter (SF). A quarter wave-plate (QWP) placed before or after the DMD directly generates the vector modes in the circular polarisation basis.}
\label{setup}
\end{figure}

Our proposal to generate arbitrary vector light fields relies on the fact that DMDs can modulate any polarisation state and therefore can tailor simultaneously, polarisation, phase and amplitude. To better understand our approach, Fig. \ref{setup} shows a schematic representation of our device. Here, an expanded and collimated (by lenses L$_1$ and L$_2$) diagonally polarised laser beam, $ \vec{u}_0(x,y)=u_0(x,y)( \hat{h}+\hat{v})/\sqrt{2}$ is separated into its vertical and horizontal polarisation components by a Wollaston prism (WP). A $4f$ imaging system composed by lenses L$_3$ and L$_4$ redirects both polarisation components towards the centre of a DMD (DLP Light Crafter 6500 from Texas Instruments), where they impinge under slightly different angles (separated by $\approx 1.5^\circ$) but are spatially overlapped. The DMD displays a multiplexed hologram, consisting of two independent holograms corresponding to the desired spatial wave functions of each polarisation component, each superimposed with a linear diffraction grating. The grating period of the holograms, in combination with the different input angles, are carefully selected such that the first diffraction order of each beam overlaps with each other along a common propagation axis where the desired complex vector field $\vec{u}(\rho,\varphi)$ is generated. The desired vectorial light fields are selected by placing a spatial filter (SF) in the far field plane of a telescope imaging the DMD plane, realised with lenses L$_5$ and L$_6$. In addition, a quarter wave plate (QWP) placed before or after the DMD directly generates the vector field in the circular polarisation basis $\hat{l}, \hat{r}$. Notably, our device can generate arbitrary vector fields at high speed rates and without any mechanical movement of optical components by simply adjusting the digital holograms displayed on the DMD.

To demonstrate our technique and without loss of generality, we will restrict our analysis to cylindrical vector modes, which can be represented using the circular polarisation basis ($\hat{r}$, $\hat{l}$) and the Laguerre-Gaussian ($LG_p^\ell$) modes of the cylindrical coordinates ($\rho$,$\varphi$) as \cite{Galvez2012,Chen2014}, 
\begin{equation}
\vec{u}(\rho,\varphi) =  \cos \left(\theta\right) LG_{p_1}^{\ell_1}(\rho,\varphi){\text e}^{i\alpha} \hat{r}+ \sin \left(\theta\right) LG_{p_2}^{\ell_2}(\rho,\varphi){\text e}^{-i\alpha} \hat{l}.
\label{Vectormodes}
\end{equation}
\noindent 
Here, $\ell_i \in \mathbb{Z}$ and $p_i \in \mathbb{N}$ are the azimuthal and radial indices, respectively, of the Laguerre-Gaussian modes. The former, known as the topological charge, is associated to the number of times the phase wraps around the optical axis, where an intensity null for $\ell\neq0$ is exhibited. The latter is associated to the number of intensity maximums ($p+1$) along the radial direction. The unit vectors $\hat{r}$ and $\hat{l}$  denote the right- and left- handed polarisation basis. The coefficients $\cos\theta$ and $\sin\theta$ ($\theta\in[0,\pi/2]$) are weighting factors that allow a smooth transition of the field $\vec{u}(\rho,\varphi)$, from purely scalar ($\theta=0$ and $\theta=\pi/2$) to vector ($\theta=\pi/4$) \cite{Selyem2019}. In addition, the term ${\text e}^{i\alpha}$ ($\alpha\in[-\pi/2,\pi/2]$) generates a phase difference between both polarisation components. For the sake of simplicity in what follows we will omit the explicit dependence of $LG_p^\ell$ on $(\rho,\varphi)$.

\begin{figure}[h!]
\centering
\includegraphics[width=.9\textwidth]{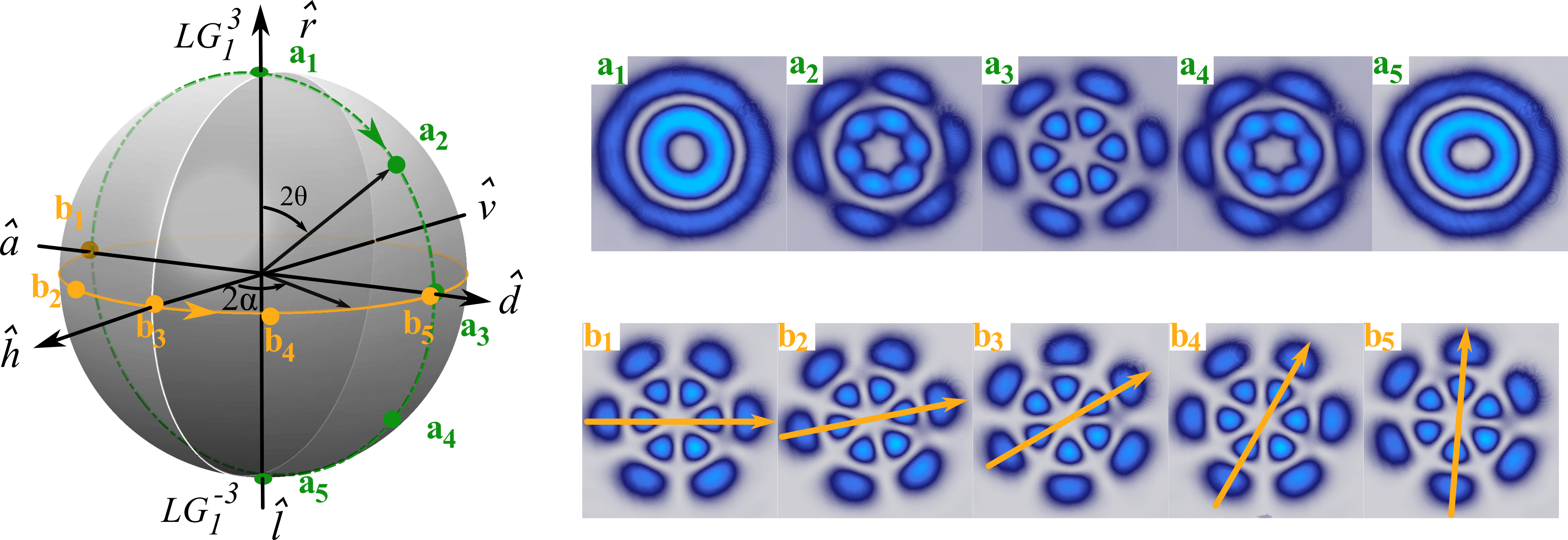}
 \caption{ Geometric representation of complex light fields on a High Order Poincar\'e Sphere (HOPS). The top right insets show experimental intensity profiles of a set of modes generated along the green dashed line (see also \blue{Media 1}). The bottom right insets show the intensity profile of beams generated along the yellow solid line (see also \blue{Media 2}). In all cases we show the shape of the vector field after passing through a linear polariser. $\hat{h}$, $\hat{v}$, $\hat{d}$, $\hat{a}$, $\hat{r}$ and $\hat{l}$ represent the horizontal, vertical, diagonal, antidiagonal, right- and left-handed unitary polarisation vectors.}
\label{Poincare}
\end{figure}

Our technique is capable to generate arbitrary modes on the High Order Poincar\'e Sphere (HOPS). As it is well known, in this representation vector beams, as defined by Eq. \ref{Vectormodes}, are mapped to unique positions ($2\alpha$,$2\theta$) on the surface of a unitary sphere constructed by assigning the north and south poles to the scalar modes $LG_{p_1}^{\ell_1}\hat{r}$ and $LG_{p_2}^{\ell_2}\hat{l}$, respectively. In this way, points along the equator correspond to pure vector beams while the remaining ones represent vector modes with elliptical polarisation. Fig. \ref{Poincare} shows representative examples of vector modes generated with our device, represented on the HOPS. For these examples we used $LG_1^{3}\hat{r}$ and $LG_1^{-3}\hat{l}$. The top-right insets of Fig. \ref{Poincare} show the intensity profile of modes generated along the green-dashed line that connects the North and South poles (see \blue{Media 1}), which were generated according to Eq. \ref{Vectormodes} by keeping $\alpha$ constant while varying $\theta\in [0,\pi/2]$. The bottom-right insets shows few examples of the modes generated along the equatorial yellow-solid line generated by keeping $\theta$ constant while changing $\alpha\in[-\pi/2,\pi/2]$  (See also \blue{Media 2}). The intensity profiles were obtained by passing the vector beams through a linear polariser oriented at an arbitrary angle.

\section{Characterisation of vector beams}

\subsection{Qualitative characterisation through Stokes polarimetry}
\begin{figure}[h]
\centering
\includegraphics[width=.9\textwidth]{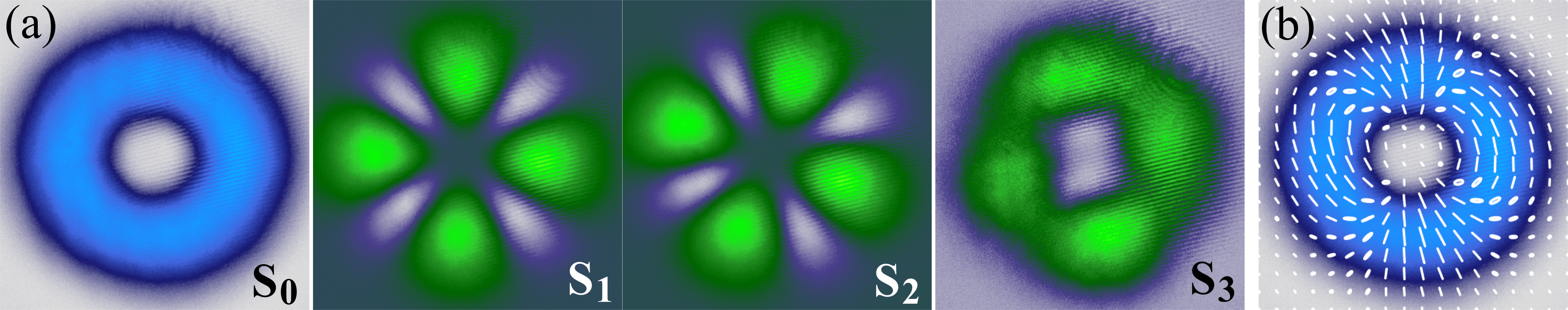}
\caption{Experimental reconstruction of polarisation using Stokes polarimetry. (a) Example of the stokes parameters S$_0$, S$_1$, S$_2$ and S$_3$ used to reconstruct the polarisation distribution of the vector mode $\frac{1}{\sqrt{2}}\left(LG_0^{-2} \hat{r}+ LG_{0}^{2}\hat{l}\right)$ shown in (b), where the local polarisation is indicated on a 18x18 grid using polarisation ellipses.}
\label{stokes}
\end{figure}

In order to quantify the capabilities of our generation technique, we programmed on the DMD various Poincar\'e beams obtained from different combinations of $p_1$, $\ell_1$, $p_2$, $\ell_2$ \cite{Galvez2012} and reconstructed their transverse polarisation distribution through Stokes polarimetry. The Stokes parameters were measured from a set of four intensities as \cite{Zhaobo2019},
\begin{align}
\centering
     S_{0}=I_{0},\hspace{5mm} S_{1}=2I_{h}-S_{0},\hspace{5mm} S_{2}=2I_{d}-S_{0},\hspace{5mm} \textrm{and}\hspace{5mm} S_{3}=2I_{r}-S_{0},
    \label{Stokes}
\end{align}
where $I_0$ is the total intensity of the beam and $I_h$, $I_d$ and $I_r$ represent the intensity of the horizontal, diagonal and right-handed polarisation components, respectively. To measure $I_h$ and $I_d$ we passed the generated complex light field $\vec{u}(\rho,\varphi)$ through a linear polarizer at $\theta=0^\circ$ and $\theta=45^\circ$, respectively. The intensity of the $I_r$ polarisation component was measured by combining a QWP at $\beta=45^0$ and a linear polarizer at $\theta=90^0$. Figure \ref{stokes} (a) shows the Stokes parameters for the case $\frac{1}{\sqrt{2}}\left(LG_0^{-2} \hat{r}+ LG_{0}^{2}\hat{l}\right)$, where $\alpha=0$, with the reconstructed polarisation distribution shown in Fig. \ref{stokes} (b) 

\begin{figure}[h]
\centering
\includegraphics[width=\textwidth]{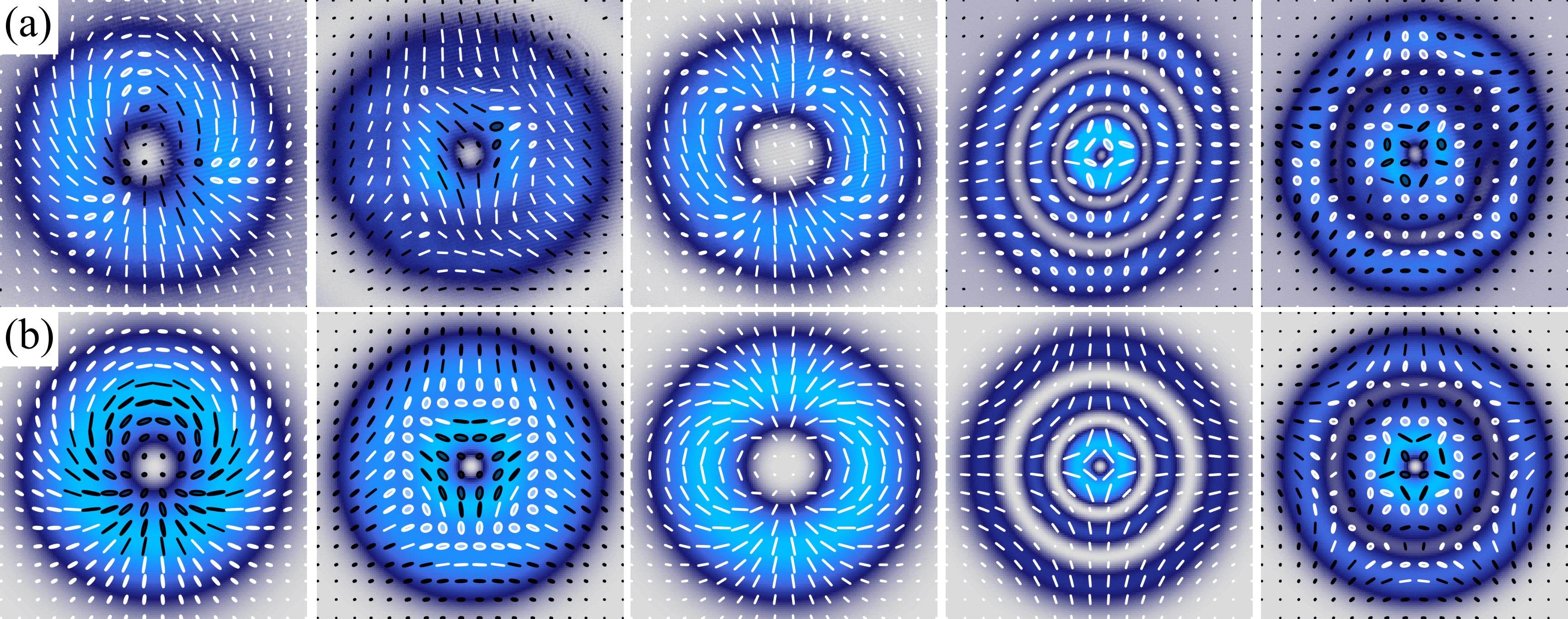}
\caption{Experimental (a) and theoretical (b) reconstruction of the transverse polarisation distribution of a set of vector modes given by pairs of scalar $LG_p^\ell$ modes, from left to right, $\left(LG_0^{-2}, LG_{0}^{+1}\right)$, $\left(LG_0^{+2},LG_{1}^{+1}\right)$, $\left(LG_0^{-2},LG_{0}^{+2}\right)$, $\left(LG_2^{+1},LG_{2}^{-1}\right)$ and $\left(LG_1^{+2},LG_{2}^{-1}\right)$.}
\label{LGvectormodes}
\end{figure}

In Fig. \ref{LGvectormodes}(a) we show a representative set of vector modes defined by the pairs of scalar $LG_p^\ell$ modes, from right to left, $\left(LG_0^{-2}, LG_{0}^{+1}\right)$, $\left(LG_0^{+2},LG_{1}^{+1}\right)$, $\left(LG_0^{-2},LG_{0}^{+2}\right)$, $\left(LG_2^{+1},LG_{2}^{-1}\right)$ and $\left(LG_1^{+2},LG_{2}^{-1}\right)$. These modes are compared to their theoretical counterpart shown in Fig. \ref{LGvectormodes}(b). 

\subsection{Quantitative characterisation through the concurrence}

As stated earlier, our device can generate any complex mode on the HOPS, from vector to scalar, as such, in this section we provide quantitative measurements of the accuracy of our technique. To this end, we will use a well-known technique that exploits the similarities between classical and quantum local entanglement, concurrence, a  measure of quantum entanglement in two dimensions. Concurrence has been identified as a proper tool to measure the degree of non-separability of vector beams, which has been termed Vector Quality Factor (VQF). The VQF assigns values between 0 and 1 to the degree of coupling between the spatial and polarisation DoF, 0 for scalar and 1 for vector modes \cite{McLaren2015,Ndagano2016}. This technique comprises the projection of the vector mode onto one DoF, polarisation in our case, which is later passed through a series of phase filters that performs a projection on the spatial DoF.  Explicitly, the VQF is determined as \cite{Ndagano2015}
\begin{equation}
\text{VQF} =\text{Re}(C)=\text{Re}\left(\sqrt{1-s^2}\right),
\label{eq:VQF}
\end{equation} 
where $s$ is the length of the Bloch vector given by
\begin{equation}
s = \left(\sum_i \langle\sigma_i\rangle^2\right)^{1/2},
\label{eq:Blochvector}
\end{equation}
and $\langle\sigma_1\rangle$, $\langle\sigma_2\rangle$ and $\langle\sigma_3\rangle$ are the expectation values of the Pauli operators. To measure this value experimentally, the two circular polarisation components are first split along different propagation paths using a polarisation grating and projected afterwards onto a set of six phase holograms that performs the projection onto the spatial DoF. The holograms, encoded on an SLM, consist of helical phases, two with topological charges $\ell$ and $-\ell$, which we represent as $R$ and $L$, respectively, and four that are superposition of the same, namely, $\exp(\text{i}\ell\phi)+ \exp(\text{i}\gamma)\exp(-\text{i}\ell\phi)$ with $\gamma = \{0, \pi/2, \pi, 3\pi/2\}$, which we represent as $H$, $D$, $V$ and $A$, respectively. The 12 intensities $\textrm{I}_{ij}$ are then measured as the on-axis values of the far field intensity recorded on a CCD camera. For the sake of clarity, the 12 required intensity measurements are explicitly shown in table \ref{Tomographic1}.

\begin{figure}[tb]
\centering
\includegraphics[width=\textwidth]{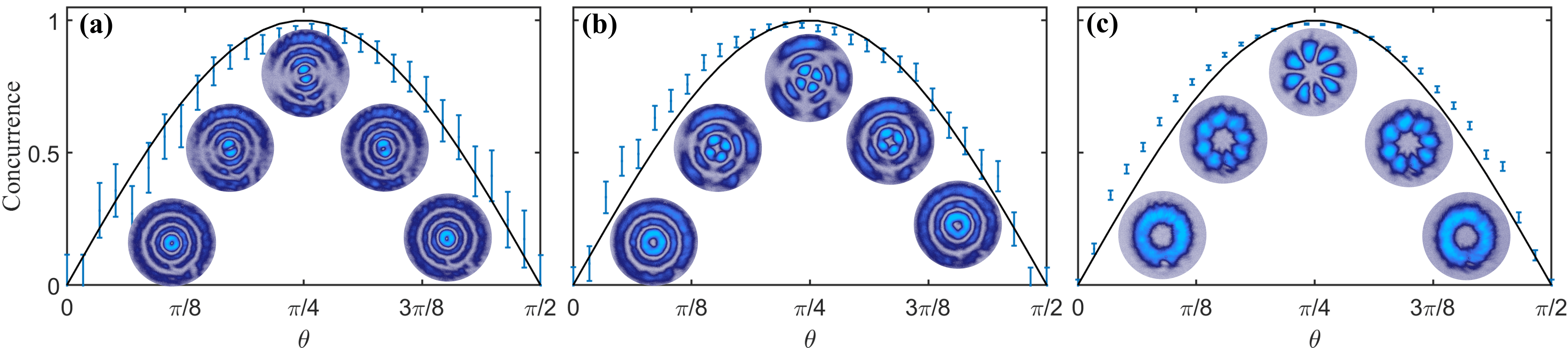}
\caption{Experimental verification of the degree of nonseparability, of the vector modes (a) $\frac{1}{\sqrt{2}}\left(LG_3^{-1} \hat{r}+ LG_{3}^{1}\hat{l}\right)$, (b) $\frac{1}{\sqrt{2}}\left(LG_2^{-2} \hat{r}+ LG_{2}^{2}\hat{l}\right)$ and (c) $\frac{1}{\sqrt{2}}\left(LG_0^{-3} \hat{r}+ LG_{0}^{3}\hat{l}\right)$. The insets show the experimental intensity profiles of the beams after a linear polariser for different degrees of non-separability.}
\label{concurrenceVQF}
\end{figure}

\begin{table}[h!] 
\centering
\setlength{\tabcolsep}{7pt}
\renewcommand*{\arraystretch}{2}
 \caption{Normalised intensity measurements $I_{ij}$ to determine the expectation values $\langle \sigma_i \rangle$. \label{Tomographic1}}
 \begin{tabular}{c|c c c c c c c}
Basis states & $ \ell^+$  & $\ell^-$  & $ \gamma_0 $ & $ \gamma_{\pi/2}$ & $ \gamma_\pi $ & $ \gamma_{3\pi/2}$ & \\
\hline \hline
$ r$ & $I_{r \ell^+}$ & $I_{r \ell^-}$ & $I_{r \gamma_0}$ & $I_{r \gamma_{\pi/2} }$&  $I_{r \gamma_\pi}$ & $I_{r \gamma_{3\pi/2}}$  &\\ 
$ l$ & $I_{l \ell^+}$ & $I_{l \ell^-}$ & $I_{l \gamma_0}$ & $I_{l \gamma_{\pi/2} }$& $I_{l \gamma_\pi }$ & $I_{l \gamma_{3\pi/2}}$& \\
\end{tabular}
\end{table}
Here, for example, $I_{r\ell^+}$ represents the intensity of the right circular polarisation component after its projection on the $\exp(\text{i}\ell\phi)$ phase filter. Finally, the expectation values are explicitly computed from the twelve intensity measurement $\textrm{I}_{ij}$ as, 
\begin{align}
\nonumber
&\langle \sigma_1\rangle = (\textrm{I}_{r\gamma_0}+\textrm{I}_{l\gamma_0})-(\textrm{I}_{r\gamma_\pi}+\textrm{I}_{l\gamma_\pi}),\\
&\langle \sigma_2\rangle = (\textrm{I}_{r\gamma_{\pi/2}}+\textrm{I}_{l\gamma_{\pi/2}})-(\textrm{I}_{r\gamma_{3\pi/2}}+\textrm{I}_{l\gamma_{3\pi/2}}),\\
\nonumber
&\langle \sigma_3\rangle = (\textrm{I}_{r\ell^+}+\textrm{I}_{l\ell^+})-(\textrm{I}_{r\ell^-}+\textrm{I}_{l\ell^-}).
\end{align}
Figure \ref{concurrenceVQF} shows some of our experimental results of the concurrence for three different vector modes generated with our device. The specific modes presented here are, $\frac{1}{\sqrt{2}}\left(LG_3^{-1} \hat{r}+ LG_{3}^{1}\hat{l}\right)$, shown in Fig. \ref{concurrenceVQF}(a) $\frac{1}{\sqrt{2}}\left(LG_2^{-2} \hat{r}+ LG_{2}^{2}\hat{l}\right)$, shown in Fig. \ref{concurrenceVQF}(b) and $\frac{1}{\sqrt{2}}\left(LG_0^{-3} \hat{r}+ LG_{0}^{3}\hat{l}\right)$, shown in Fig. \ref{concurrenceVQF}(c). We changed digitally the amplitude coefficients of each mode, namely, $\cos(\theta)$ for $\theta\in[0,\pi/2]$, to observe their transition from scalar to pure vector modes. The insets of each plot show the intensity distribution of the input field $u(\vec{r})$, as viewed on a CCD camera after a linear polarizer. Small intensity fluctuations at the detector caused uncertainty in the measured concurrence (shown as error bars in Fig. \ref{concurrenceVQF}) which was characterised for each intensity measurement by taking the standard deviation after averaging over the 64 central pixels, as detailed in ref.~\cite{Selyem2019}. 


Recently we proposed an alternative method to measure the concurrence, which does not require the projection on the spatial DoF. This technique is based on Stokes projections, providing a basis-independent measurement. For this, the Stokes parameters of the generated beams are calculated by projecting the vector beams onto the polarisation basis vectors $\hat{h}$, $\hat{v}$, $\hat{a}$, $\hat{d}$, $\hat{r}$ and $\hat{l}$ using a quarter-wave plate and a linear polariser. The concurrence in this case is computed as,
\begin{equation}
C=\sqrt{1-\left(\frac{h-v}{h+v} \right)^2-\left(\frac{d-a}{d+a} \right)^2-\left(\frac{r-l}{r+l} \right)^2},
\end{equation}
where \textit{h, v, a, d, r} and $l$ are the projections described above, integrated over the beam profile. A complete explanation of this technique as well as experimental measurements are given in \cite{Selyem2019}.


\section{Conclusions}
In this manuscript we put forward a novel technique to generate arbitrary complex light vector fields based on a Digital Micromirror Devices (DMDs). DMDs have been around for several decades but it is only in the last decade that they became an alternative device for the generation of structured light beams. Nonetheless the polarisation-independence of DMDs has gone almost unnoticed, as such, in our approach we take full advantage of this property. For this, we illuminate the DMD with two beams of orthogonal polarisation impinging at different angles so that the first diffraction order of the beams overlap with each other. To this end, we multiplexed on the DMD two holograms with unique spatial frequencies that modulate each beam's amplitude and phase independently so that the beam emerging from the DMD results in a complex vector light field. Importantly, our technique can generate any vector light field with arbitrary degrees of non-separability and any polarisation distribution. To demonstrate this, we characterised the modes generated with our device using both, a qualitative and a quantitative measurement. The first through a Stokes polarimetry reconstruction whereas the later with a measure of the degree of non-separability between the spatial and polarisation DoF, concurrence. Both measurements show excellent correlation between experimental measurements and the predicted theory. Our device can generate vector beams with a choice of VQF within +-5\% of the theoretical VQF when beams are close to fully vectorial, and presumably performs just as well at low VQF, but our measurement technique has low signal-to-noise ratio there.

\section*{Funding}
This work was partially supported by the National Nature Science Foundation of China (NSFC) under Grant No. 61975047
\section*{Disclosures}
The authors declare that there are no conflicts of interest related to this article.
\bibliography{References}
\end{document}